\documentclass[aps,twocolumn,preprintnumbers]{revtex4}

\usepackage{graphicx}  
\usepackage{subfigure}
\usepackage{multirow}

\usepackage{fancyhdr}
\usepackage{longtable}
\usepackage{parskip}
\usepackage[T1]{fontenc}
\usepackage{dcolumn}   

\usepackage{bm}        
\usepackage{amsfonts}  
\usepackage{amsmath}   
\usepackage{amssymb}   

\newcommand{\pwisein}{\left\{ \begin{array}{ll}}
\newcommand{\pwiseout}{\end{array}\right.}

\begin{document}

\title{$q$-Index Degree Distribution in Random Networks via Superstatistics}

\author{Huilin Wang${^1}$}
\affiliation{$^{1}$Key Laboratory of Quark and Lepton Physics (MOE) and Institute of Particle Physics, Central China Normal University, Wuhan 430079, China}

\author{Weibing Deng$^{1}$}
\email{wdeng@mail.ccnu.edu.cn}
\affiliation{$^{1}$Key Laboratory of Quark and Lepton Physics (MOE) and Institute of Particle Physics, Central China Normal University, Wuhan 430079, China}

\author{Zhekai Chen$^{1}$}
\affiliation{$^{1}$Key Laboratory of Quark and Lepton Physics (MOE) and Institute of Particle Physics, Central China Normal University, Wuhan 430079, China}


\begin{abstract}  

In this study, we employ a superstatistical approach to construct $q$-exponential and $q$-Maxwell-Boltzmann complex networks, generalizing the concept of scale-free networks. By adjusting the crossover parameter $\lambda$, we control the degree of the $q$-exponential plateau at low node degrees, allowing a smooth transition to pure power-law degree distributions. Similarly, the parameter $b$ modulates the $q$-Maxwell-Boltzmann curvature, facilitating a shift toward pure power-law networks. This framework introduces a novel perspective for constructing and analyzing scale-free networks. Our results show that these additional degrees of freedom significantly enhance the flexibility of both network types in terms of topological and transport properties, including clustering coefficients, small-world characteristics, and resilience to attacks. Future research will focus on exploring the dynamic properties of these networks, offering promising directions for further investigation.
\end{abstract}

\maketitle 

\section{Introduction}


In nonequilibrium statistical mechanics, superstatistical models serve as a robust framework for analyzing complex systems subjected to significant environmental changes and temperature fluctuations\cite{beck2003superstatistics} A superstatistical complex system is mathematically characterized by the integration of multiple statistical distributions\cite{albert2002statistical}, one representing equilibrium statistical mechanics and the other reflecting a gradually varying system parameter. Central to this approach is the requirement for a significant separation of timescales: the local relaxation time of the system should be substantially shorter than the typical timescale of change\cite{beck2005time}. The superstatistical framework has found applications across various complex systems, including hydrodynamic turbulence \cite{beck2007statistics},frequency fluctuations in power grids\cite{schafer2018non},Application to the SYM-H geomagnetic index\cite{sanchez2024testing}, analyze complex network formation from random graph fluctuations\cite{abe2006hierarchical}. and air pollution statistics \cite{williams2020superstatistical}.

In past research, the analysis of complex networks has primarily focused on network growth models with preferential attachment, often overlooking those without preferential attachment mechanisms. These models tend to result in different degree distribution forms and network characteristics\cite{sampaio2023random}. While these studies have made significant progress in certain areas, they have failed to fully explain the diversity and heterogeneity of real-world complex networks, especially in the context of nonequilibrium dynamics.To address this gap, we propose a model based on superstatistics, drawing inspiration from the theory of Brownian motion in nonequilibrium physics \cite{carro2016coupled,thurner2005nonextensive,wedemann2009nonextensivity}. The goal is to reveal the statistical features of complex network structures across different scales\cite{albert2002statistical}. Specifically, our study explores how concepts such as the q-exponential distribution, the q-Maxwell–Boltzmann distribution, and power-law distributions intertwine and jointly shape the evolution of networks at both local and global scales.

Our work not only fills a critical gap in network theory but also provides new insights into understanding self-organization and nonlinear phenomena within complex systems. This approach offers a powerful theoretical tool for future tasks in network optimization and dynamic prediction



\section{Theory}
In Brownian motion, we observe the trajectory of a Brownian particle in an environment that is constantly changing. The particle experiences relatively fast dynamics due to speed, while changes in ambient temperature act as a slow driving factor for its motion\cite{beck2003superstatistics}. The temperature affects the moving system, and the particle's speed follows this influence.:
\begin{equation}\label{eq1}
v= - \gamma \dot{v}+ \sigma L\left ( t\right )
\end{equation}

In this model, the parameter $\beta$ is not constant but changes over time on the scale T and over space on the scale L. These changes are a result of the complex dynamics of the environment of the Brownian particle. It has been demonstrated that when averaging over the fluctuating $\beta$, this generalized Langevin model produces Tsallis statistics for v if $\beta$ is a $\chi^2$-distributed random variable. Additionally, the distributions obtained for v were found to accurately match distributions of longitudinal velocity differences in turbulent Taylor-Couette flows\cite{beck2001dynamical,beck2001measuring}, as well as measurements in Lagrangian turbulence\cite{beck2001small,la2001fluid}.

Let's extend this approach to general distributions $f(\beta)$ and general (effective) Hamiltonians. In the long-term (when $t\gg T$), the stationary probability density of our nonequilibrium system is determined by Boltzmann factors ${e}^{-\beta E}$ associated with the cells, which are averaged over the various fluctuating inverse temperatures $\beta$. If $E$ is the energy of a microstate associated with each cell, we can express this as:

\begin{equation}\label{eq1}
B(E)=\int_{0}^{\infty }f(\beta){e}^{-\beta E}d\beta 
\end{equation}

In our nonequilibrium system, the parameter B represents the superstatistics of the system and can be thought of as an effective Boltzmann factor. It describes the statistics of the statistics (${e}^{-\beta E}$) of the system's cells. B(E) may differ significantly from the ordinary Boltzmann factor, which is obtained when $f(\beta )=\delta (\beta -{\beta }_{0})$. For example, in the case of a Brownian test particle with mass 1, the energy E is given by $E=\frac{1}{2}{v}^{2}$. In this case, the long-term stationary state is a combination of Gaussian distributions ${e}^{-\beta E}$ weighted by the probability density $f(\beta)$, which represents the likelihood of observing a certain $\beta$. It's important to note that our analysis applies to arbitrary energies E associated with the cells, not just $E=\frac{1}{2}{v}^{2}$. The central hypothesis of our paper is that generalized Boltzmann factors of the form (2) are physically relevant for a wide range of dynamically complex systems with fluctuations.

In contrast to previous studies that used preferential attachment models to create networks with degree distributions resembling q-exponential behavior, we use the configuration model\cite{molloy1995critical,newman2001random} to construct our random networks. This approach ensures that specific topological properties, such as inherent node correlations, are not a result of the growth process\cite{soares2005preferential,brito2016role,brito2019scaling,ochiai2009construction}, but rather directly linked to the q-exponential shape of the degree distribution.

 We begin by assigning a specific degree ${k}_{i}$ to each node i, which is chosen from a q-exponential distribution. Since the degrees are whole numbers, we randomly choose a number ${x}_{i}$ from a q-exponential distribution and define ${k}_{i}$ as the largest whole number less than ${x}_{i}$. To prevent creating lots of small disconnected clusters, we only include nodes with a degree ${k}_{i}\geqslant 2$, meaning ${k}_{min}=2$. We represent the as-yet unconnected ${k}_{i}$ degrees on node i with ${k}_{i}$ "stubs"\cite{sampaio2023random}. Then we move on to connect nodes in pairs. To do this, we select two different nodes with probabilities based on their number of stubs. If these two nodes are not yet connected, we place a link between them and reduce the number of stubs for each of the two nodes by one. This process continues until all stubs have been connected. If there are remaining stubs at the end that could not be connected, these stubs are removed, and the degree originally assigned to the corresponding nodes in the network is finite.
 \begin{figure}
\includegraphics[width=3.0in]{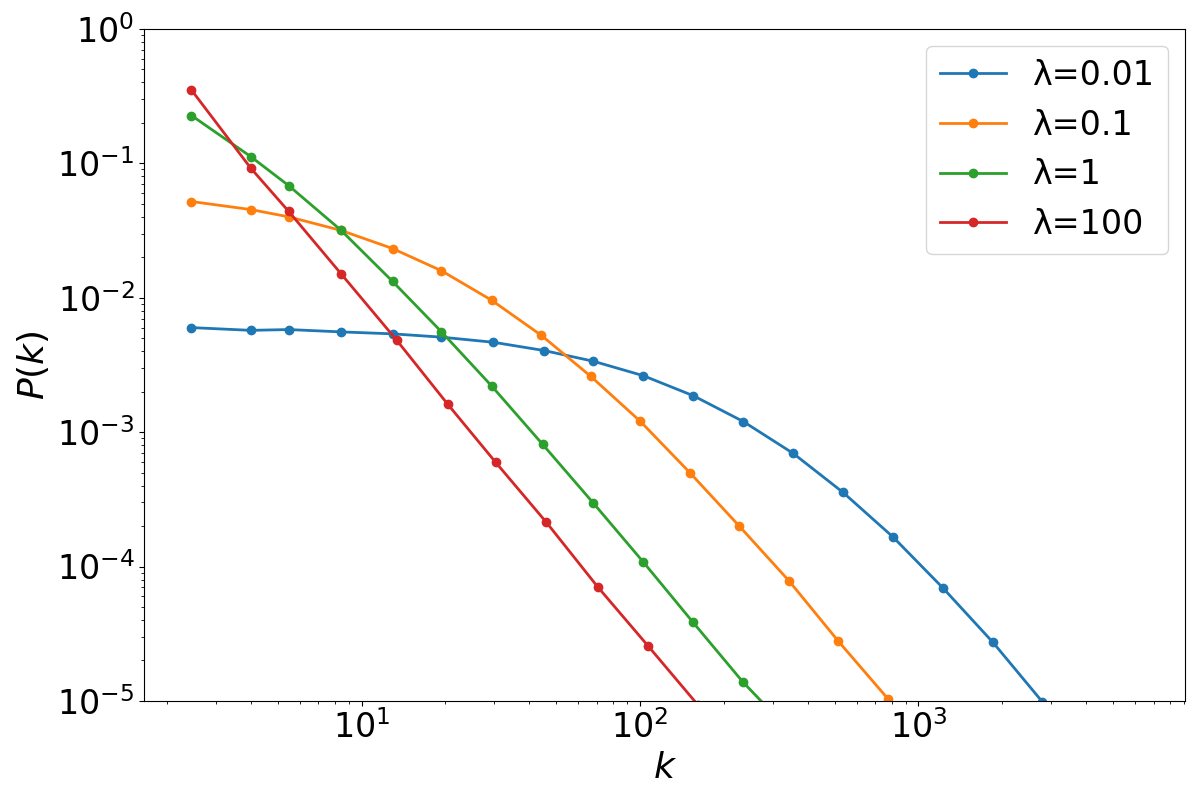}
\caption{\label{setup}
Schematic of the ball drop and the measurement.
Degree distribution of q-exponential networks (symbols) compared with expected distribution of Eq.(3) (solid lines) for q=1.4 and for $\lambda=0.01$ (blue circles), $\lambda=0.1$ (yellow circles), $\lambda=1$ (green circle), and $\lambda= 100$ (red circle). The inset shows a comparison between the degree distributions of a q-exponential with $\lambda=100$ and a pure scale-free distribution (dashed black line) with $q=1.4$ ($\gamma=2.5$). These results are obtained for networks with size $N=20000$ by averaging over 100 samples.}
\end{figure}

\begin{figure}
\includegraphics[width=3.0in]{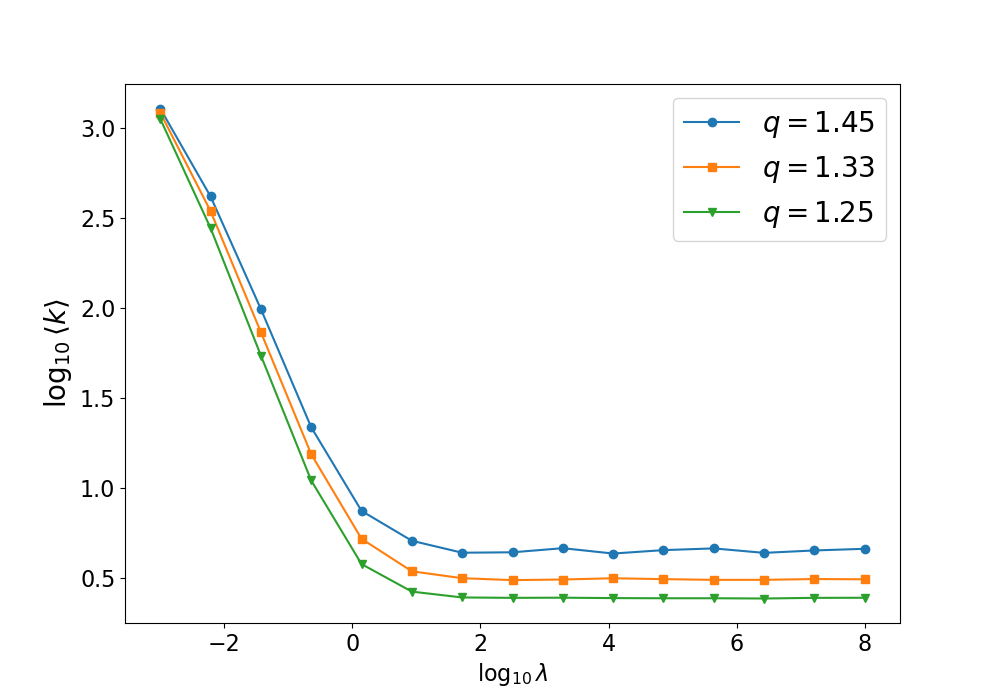}
\caption{\label{setup}
Dependence of the average degree $\left \langle k\right \rangle$ on the parameter$\lambda$ for different values of q.
These curves correspond to q=1.4,(black circles),q=1.33,(red squares), and q=1.25, (blue stars). For values of $\lambda$<1, the average degree follows $\left \langle k\right \rangle\backsim {\lambda }^{-1}$, as expected for q-exponential distributions. In the pure power-law limit ($\lambda \gg 1$), the average degree saturates at a value independent of $\lambda$.}
\end{figure}

\begin{figure}
\includegraphics[width=3.0in]{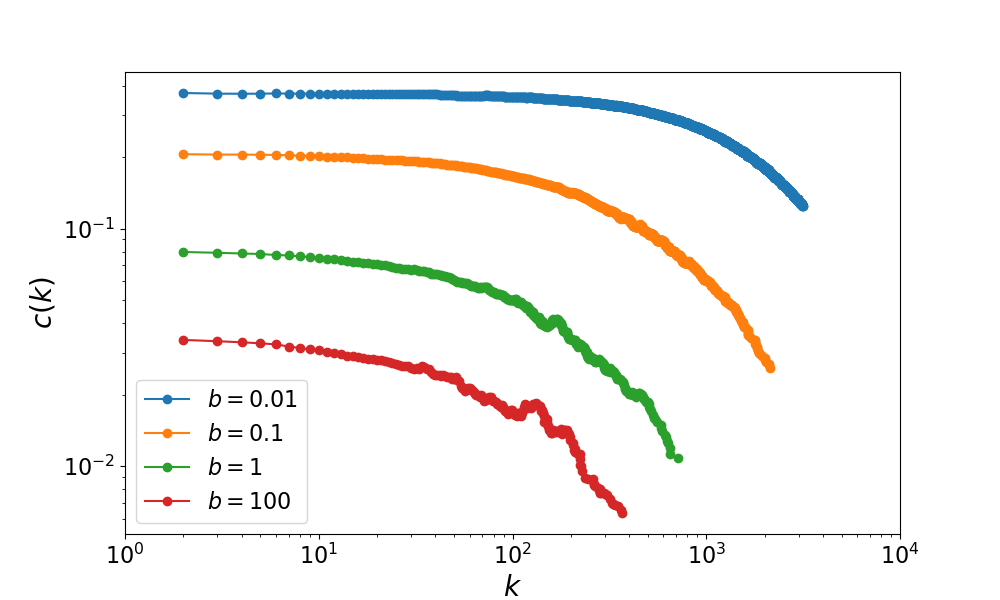}
\caption{\label{setup}
Relationship between the degree $k$ and the clustering coefficient $c(k)$ for different values of the parameter $\lambda$. The log-log plot shows the clustering coefficient $c(k)$ as a function of degree $k$ for four distinct values of $\lambda$: 0.01, 0.1, 1, and 100. It is evident that increasing $\lambda$ results in a systematically lower clustering coefficient across all degree ranges, indicating a reduction in the modularity and locality of connections within the network. This trend suggests that higher values of $\lambda$ lead to more homogeneous connectivity, reducing the prevalence of tightly clustered subnetworks.}
\end{figure}

\begin{figure}
\includegraphics[width=3.0in]{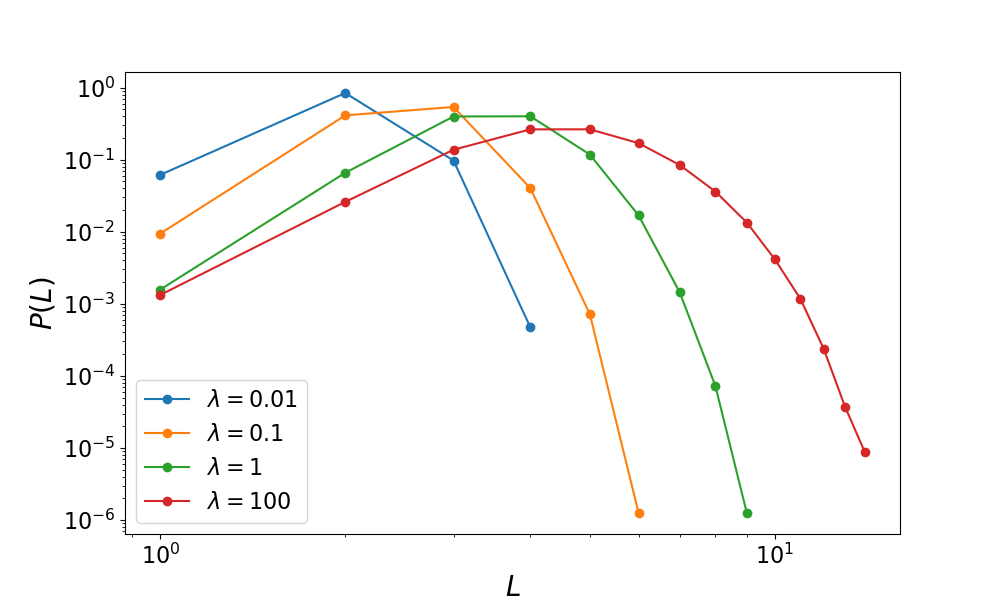}
\caption{\label{setup}
Shortest path length distribution $P(L)$ for different values of $\lambda$. The log-log plot illustrates how the distribution of shortest path lengths $P(L)$ changes as a function of $\lambda$: 0.01, 0.1, 1, and 100. As $\lambda$ increases, the distribution shifts to longer average path lengths, indicating a transition towards networks with increased global connectivity and reduced local clustering. This behavior is especially evident in the increased prevalence of longer paths for higher values of $\lambda$. The figure clearly shows that as $\lambda$ grows, the network structure becomes more homogeneous, leading to a broader distribution of path lengths.}
\end{figure}

\begin{figure}
\includegraphics[width=3.0in]{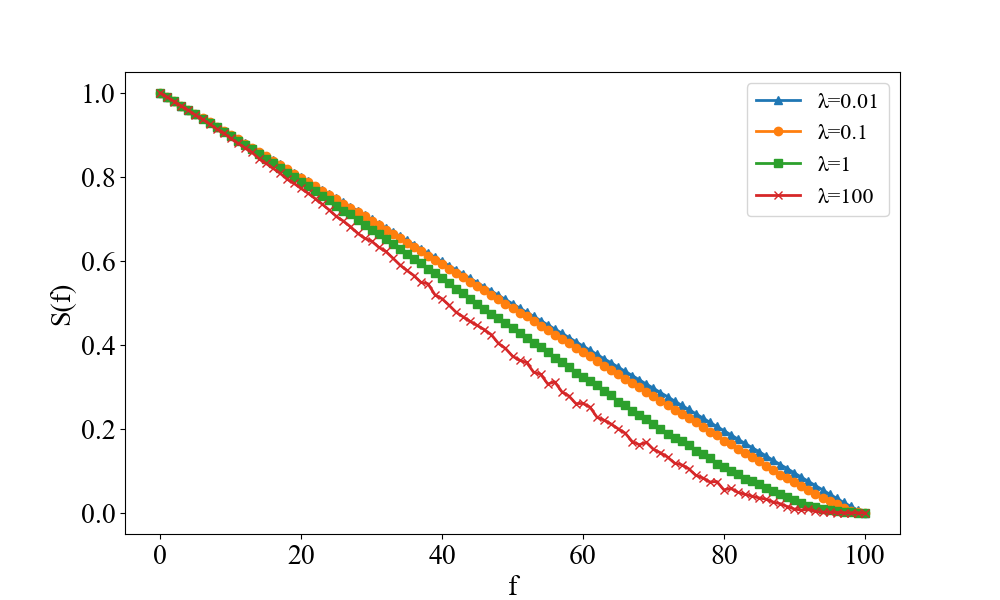}
\caption{\label{setup}
Effect of random attacks on the network structure for different values of $\lambda$. The plot shows the relative size of the largest connected component as a function of the proportion $f$ of nodes removed, for $\lambda$ values of 0.01, 0.1, 1, and 100. As $\lambda$ increases, the network becomes more susceptible to fragmentation under random attack, indicated by a steeper decline in the size of the largest connected component. The results suggest that networks with higher $\lambda$ values are less resilient to random failures, possibly due to the reduced local clustering and greater reliance on fewer high-degree nodes, which when removed, cause rapid disintegration of the global connectivity.}
\end{figure}

\begin{figure}
\includegraphics[width=3.0in]{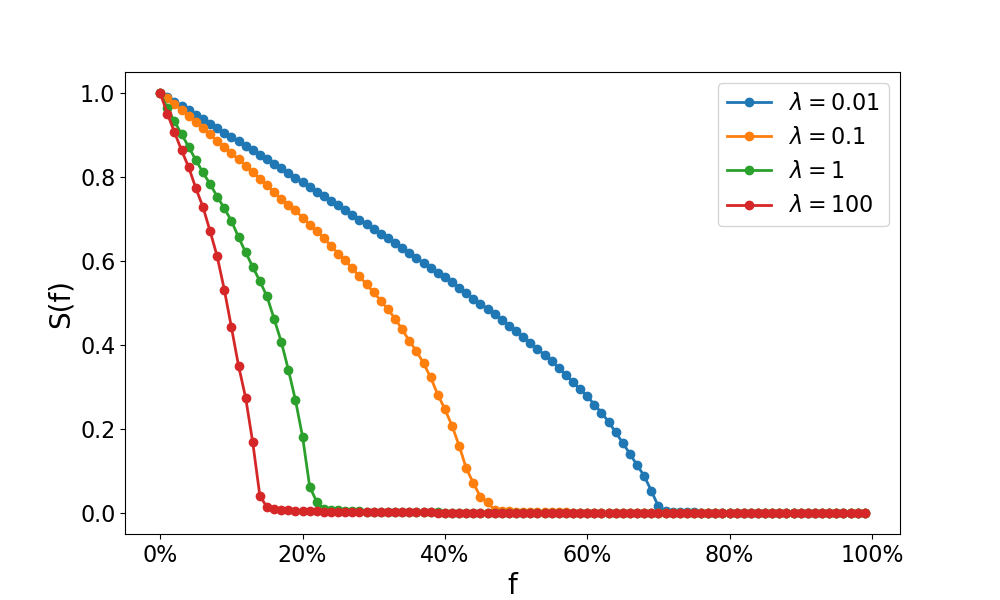}
\caption{\label{setup}
Effect of deliberate attacks on the network structure for different values of $\lambda$. The plot shows the relative size of the largest connected component as a function of the proportion of nodes removed, where nodes are deliberately targeted based on decreasing degree (i.e., high-degree nodes are removed first). The results are presented for different values of $\lambda$: 0.01, 0.1, 1, and 100. The plot reveals that networks with higher $\lambda$ values are significantly more vulnerable to targeted attacks, with the largest connected component collapsing much more rapidly compared to networks with lower $\lambda$ values. This behavior indicates that networks with higher $\lambda$ exhibit more hub-like nodes, which, when removed, lead to severe fragmentation. Conversely, lower $\lambda$ values produce more resilient structures under targeted attacks due to the more homogeneous degree distribution.}
\end{figure}

\section{q-exponential networks}


Let's consider a network that follows an exponential distribution, given by the formula $p(k/\beta)=\beta exp(-\beta k)$. If the constant $\beta$ can vary and follow a certain distribution, how will this affect the overall structure of the network? In this paper, the variation is described as a ${\chi }^{2}$ distribution\cite{williams2020superstatistical}.
\begin{equation}\label{eq2}
\begin{split}
f\left ( \beta \right )=\frac{1}{\varGamma \left ( \frac{n}{2}\right )}{\left ( \frac{n}{2{\beta }_{0}}\right )}^{\frac{n}{2}}{\beta }^{\frac{n}{2}-1}exp\left ( -\frac{n\beta }{2{\beta }_{0}}\right )
\end{split}
\end{equation}

where n represents the number of degrees of freedom and ${\beta }_{0}$ is the mean of $\beta$ when integrating out the $\beta$ parameter. The marginal distribution $p(k)$ is calculated as $p(k)=(2-q)\lambda {[1-(1-q)\lambda k]}^{-\frac{1}{(q-1)}}$, and the q-exponential distribution is given by\cite{williams2020superstatistical},

\begin{equation}\label{eq3}
\begin{split}
p(k) &= \int_{0}^{\infty} p(k/\beta) f(\beta) d\beta \\
&= p(k) = (2-q)\lambda \left[1 - (1-q)\lambda k\right]^{-\frac{1}{(q-1)}},
\end{split}
\end{equation}
has two parameters,$q\geqslant 1$ and $\lambda \geqslant 0$.$\lambda =\frac{1}{\kappa (3-2q)}$ and $-\frac{n}{2}-1=\frac{1}{1-q}$ with $\frac{1}{2}(q-1)\lambda =\frac{{\beta }_{0}}{n}$ $\kappa =\left \langle k\right \rangle$,${\beta }_{0}=\left \langle \beta \right \rangle$ number of nodes$N=20000$

In Figure 1, the degree distribution obtained by this method closely follows the expected q-exponential form of Equation (3). When the parameter $\lambda$ is much less than 1, there is a noticeable plateau for small k. Conversely, when $\lambda$ is much greater than 1, our degree distribution becomes effectively identical to a scale-free distribution with the same ${k}_{min}$. By decreasing the parameter $\lambda$, we can continuously move away from a scale-free degree distribution and widen the plateau\cite{sampaio2023random}. Therefore, varying $\lambda$ will allow us to identify the effect of the deviations from pure scale-freeness \cite{posfai2016network}that are specific to q-exponential distributions.

In Figure 2, we illustrate how the average degree $\left \langle k\right \rangle$ changes with $\lambda$ for different values of the parameter q. In line with expectations for a q-exponential network, for small values of $\lambda$, the average degree is proportional to ${\lambda }^{-1}$. However, for sufficiently large values of $\lambda$, the degree distribution transitions into a power law. In this latter scenario, the average degree becomes independent of $\lambda$. Consequently, scale-free networks have considerably more least-connected nodes compared to networks with a q-exponential degree distribution and small $\lambda$. Smaller values of $\lambda$ lead to denser networks, increasing the number and degree of their hubs. These topological differences can result in significant changes in the structural properties of complex networks, as well as in the static and dynamic behavior of models when implemented on these substrates.   



Random q-exponential networks also demonstrate small-world behavior, as depicted in Fig. 3. For smaller values of $\lambda$, the distribution is skewed towards shorter paths, whereas for higher $\lambda$, the distribution shows a heavier tail, indicating the presence of longer paths. In Fig. 4, we also illustrate how the average clustering coefficient $c(k)$ changes with node degree $k$ for different $\lambda$ values. The clustering coefficient reflects the probability that a node's neighbors are interconnected. As $\lambda$ increases from 0.01 to 1, the overall clustering decreases, particularly for high-degree nodes, suggesting that a larger $\lambda$ results in reduced local clustering in the network.
 
In practical applications, the robustness of networks against random failures is a crucial property. In Figure 5, we display the number of nodes in the largest cluster, denoted as $S(f)$, as a function of the fraction $f$ of randomly removed nodes. This is done for a network size of $N=20000$, a $q$ value of 1.25, and different values of $\lambda$. Our findings indicate that for networks generated with $\lambda \leq 0.1$ and this specific $q$ value, they are completely robust. This means that they consistently maintain a finite fraction of nodes in the largest cluster ($S(f)$) when subjected to random attacks\cite{molloy1995critical}, with a critical threshold of $f_{c}=1$.On the other hand, for $q$-exponential networks with $\lambda > 1$, there exists a threshold $f_{c} <1$, beyond which the structure is entirely disrupted, resulting in $S(f)=0$ for any $f \geq f_{c}$. It's evident that as a $q$-exponential network approaches a purely scale-free one, particularly for large values of $\lambda$, it becomes more fragile against random failures. Additionally, this vulnerability is significantly amplified with an increase in the parameter $q$.

In Figure 6, we illustrate how the size of the largest cluster, denoted as $S$, changes with the fraction $f$ of removed nodes in the case of a malicious attack for $q=1.4$ (corresponding to $\gamma=2.5$) and various values of $\lambda$. The graph shows that $q$-exponential networks become less resilient as the crossover parameter $\lambda$ increases, meaning that a larger fraction of nodes needs to be removed before reaching the critical point. This trend continues until $\lambda$ reaches a sufficiently large value, at which point the scale-free behavior of the degree distribution takes over. Consequently, the $S$ versus $f$ curve for $\lambda=100$ and the corresponding curve from networks with a pure scale-free distribution perfectly overlap.



\begin{figure}[t]
\includegraphics[width=3.0in]{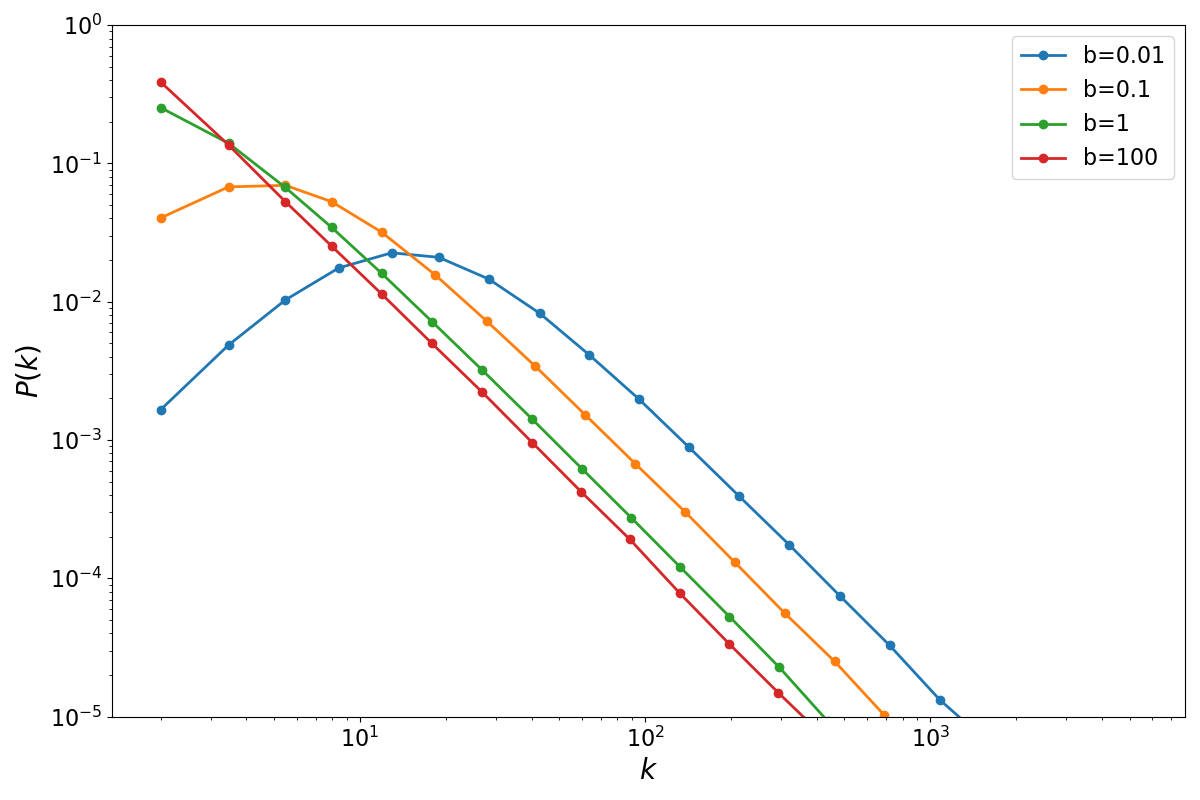}
\caption{\label{graph} Average degree distribution for $q$-Maxwell networks generated using different values of $b$ over 20 runs, with log-binning applied. The distributions exhibit different decay characteristics, indicating the effect of the parameter $b$ on the network's structural properties.
}
\end{figure}

\begin{figure}[t]
\includegraphics[width=3.0in]{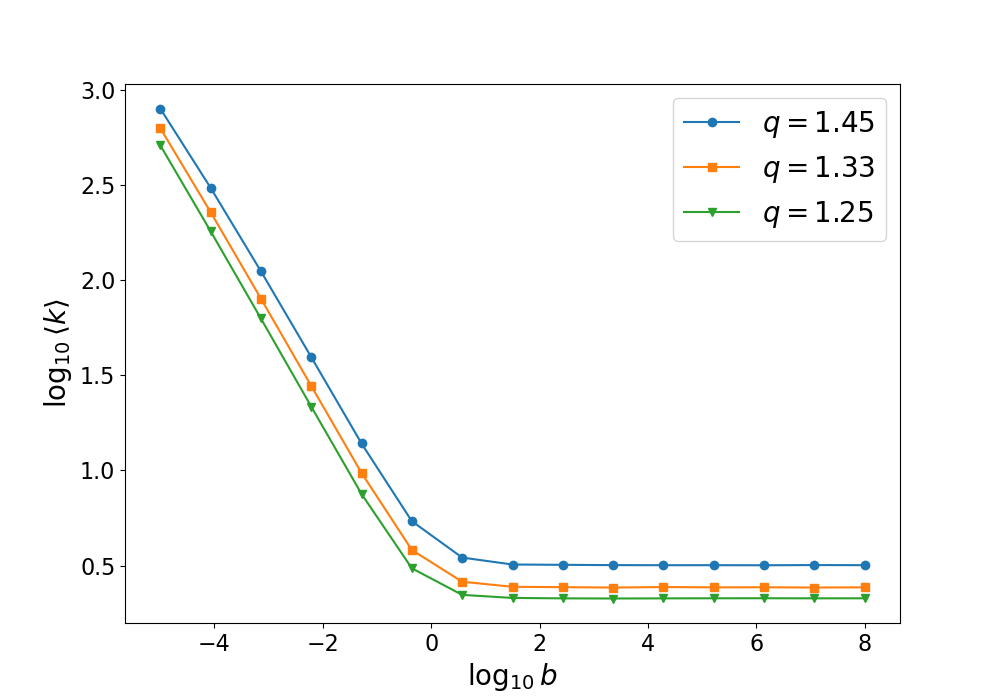}
\caption{\label{graph} Relationship between the average degree $\log_{10}(k)$ and the parameter $\log_{10}(b)$ for different values of the parameter $q$: $1.45$, $1.33$, and $1.25$. The plot shows how the average degree varies with changes in $b$, illustrating distinct trends for each value of $q$.
}
\end{figure}

\begin{figure}[t]
\includegraphics[width=3.0in]{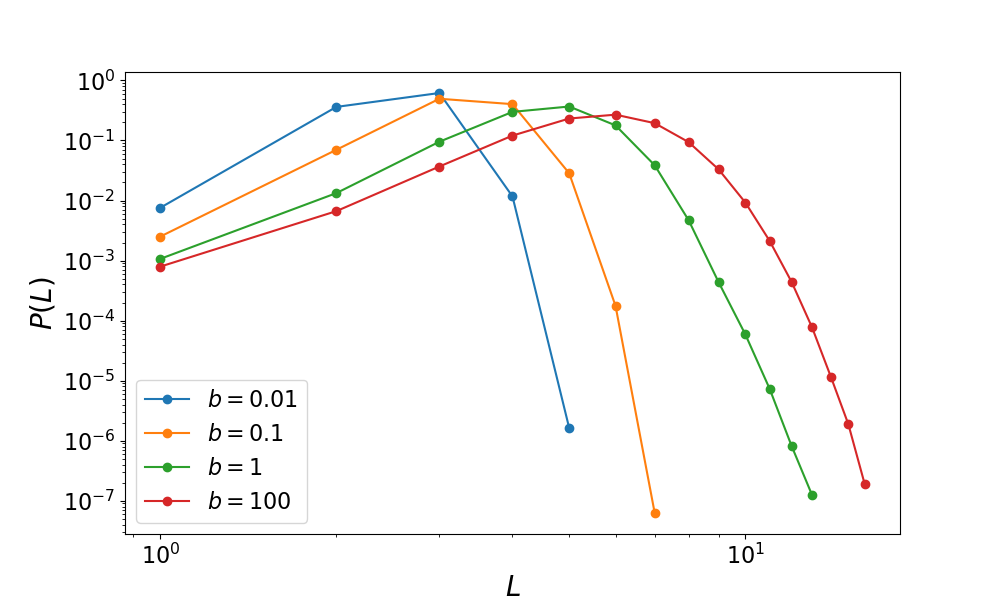}
\caption{\label{graph} Averaged shortest path length distribution $P(L)$ over 20 runs for different values of the parameter $b$: $0.01$, $0.1$, $1$, and $100$. The distribution illustrates how the average shortest path length changes with varying $b$, highlighting distinct structural characteristics of the network.
}
\end{figure}

\begin{figure}[t]
\includegraphics[width=3.0in]{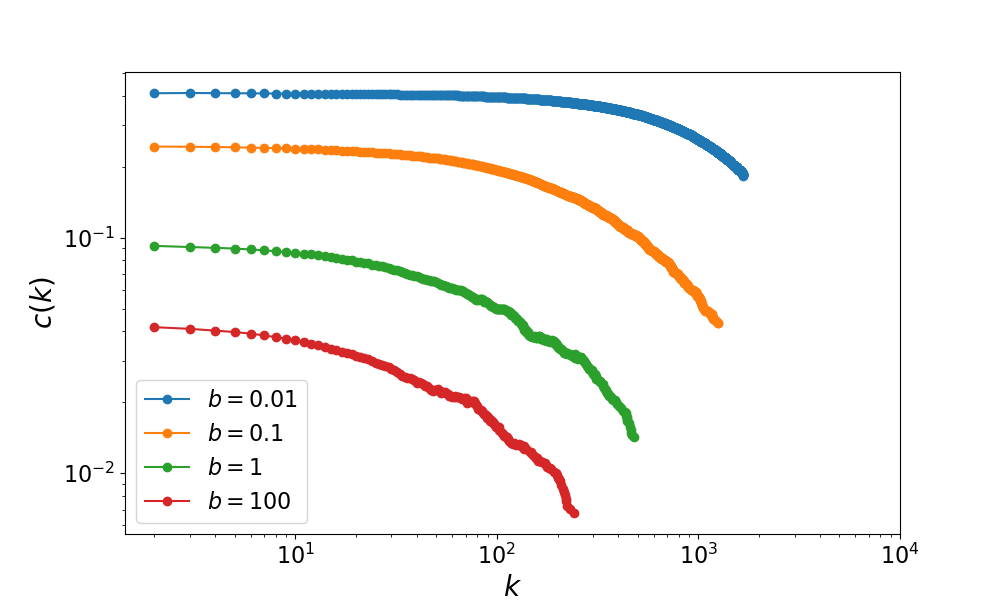}
\caption{\label{graph} Degree-clustering coefficient relationship $c(k)$ with moving average for different values of the parameter $b$: $0.01$, $0.1$, $1$, and $100$. The plot illustrates how the clustering coefficient varies with the degree $k$, highlighting the impact of the parameter $b$ on network topology.
}
\end{figure}

\begin{figure}[t]
\includegraphics[width=3.0in]{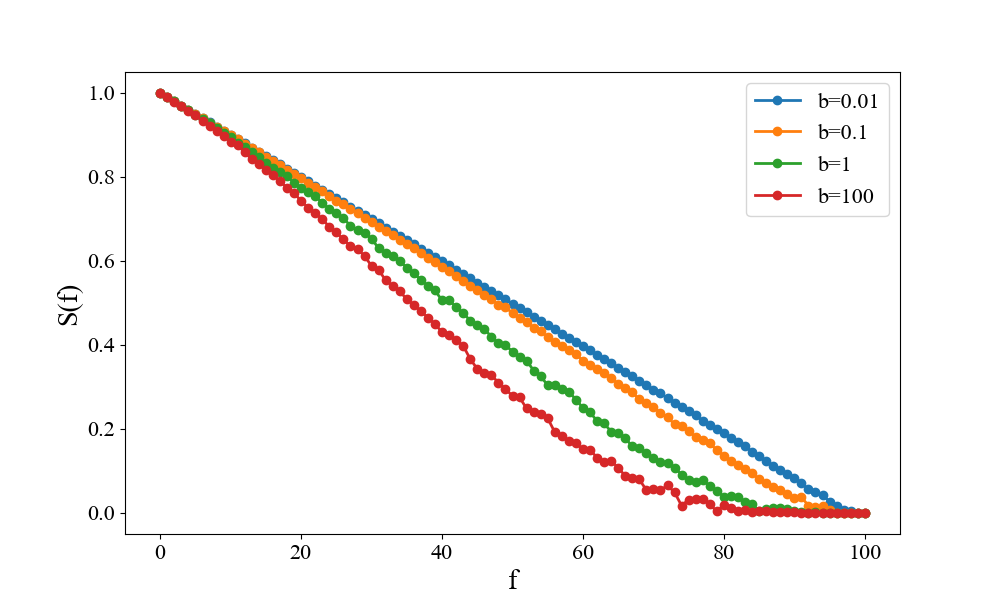}
\caption{\label{graph} The relative size of the giant connected component in the network as a function of the proportion \(f\) of nodes removed during random attacks. The results are shown for different values of the parameter \(b\): \(0.01\), \(0.1\), \(1\), and \(100\). The graph illustrates how network resilience varies with changes in \(b\).
}
\end{figure}

\begin{figure}[t]
\includegraphics[width=3.0in]{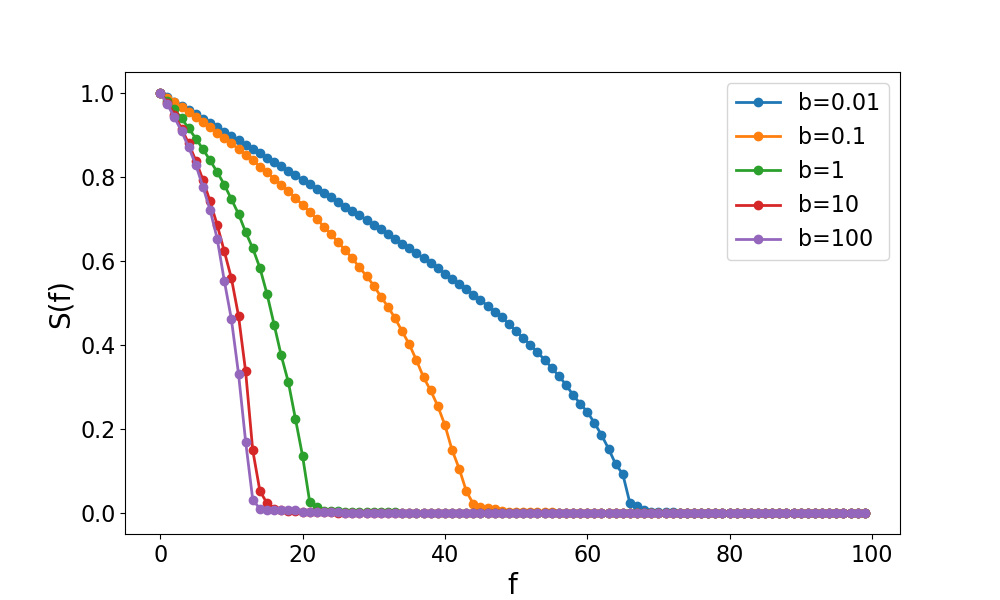}
\caption{\label{graph} Relative size of the largest connected component in the network under deliberate attacks as a function of the proportion of removed nodes. The results are shown for various values of the parameter $b$: $0.01$, $0.1$, $1$, $10$, and $100$. The graph demonstrates how network resilience is affected by deliberate node removal.
}
\end{figure}
%


\section{q-Maxwell-Boltzmann networks}

In this part, we also proposed another superstatistical model for building scale-free networks, which is given by the following expression:
\begin{equation}\label{eq2}
p(k, \sigma_{mb}) = \frac{4}{\sqrt{\pi}} \, k^2 \, \sigma_{mb}^{3/2} \, \exp\left(-\sigma_{mb} k^2\right)
\end{equation}

However, if the constant $\beta$ can vary and follow some distribution, how would the structure of the whole network change? In this case, $\beta$ follows a ${\chi}^{2}$ distribution, and the resulting probability distribution is the q-Maxwell-Boltzmann distribution\cite{williams2020superstatistical}, given by the expression:

\begin{equation}\label{eq2}
p(k) = \frac{1}{Z} \, k^{2} \, b^{\frac{3}{2}} \left[ 1 + (q - 1) b k^{2} \right]^{\frac{1}{1 - q}}
\end{equation}

where Z serves as the normalization constant ,and their are also two parameters,$1\leq q< \frac{5}{3}$ ,$b={\sigma }_{mb}$and $b> 0$.and we identify
\begin{equation}\label{eq2}
-(\frac{n+3}{2} )=\frac{1}{1-q} 
\end{equation}

\begin{equation}\label{eq2}
\frac{1}{2}(q-1)b=\frac{\beta _{0} }{n}   
\end{equation}
where $\beta _{0}$ is the mean of $\beta$,number of nodes$N=20000$
The average degree distribution of the $q$-Maxwell network model for four different values of the parameter $b$: $0.01$, $0.1$, $1$, and $100$ is shown in Figure 7. The distributions are averaged over 20 independent runs, with log-binning applied to improve clarity, particularly in the tail regions. For smaller values of $b$ (e.g., $b = 0.01$), the degree distribution decays gradually, exhibiting a heavy-tailed behavior typical of scale-free networks. As $b$ increases, the decay becomes steeper, indicating a reduced probability of observing high-degree nodes. Specifically, when $b = 100$, the distribution exhibits an exponential-like decay, suggesting a transition toward a more homogeneous structure with fewer hubs. These trends indicate that the parameter $b$ acts as a control variable, tuning the network between a heterogeneous, hub-dominated regime and a more uniform network structure.

The relationship between the average degree (in logarithmic scale) and the parameter $b$ (also in logarithmic scale) is depicted in Figure 8. The data show a sharp decrease in average degree as $b$ increases, particularly in the lower range. Initially, the average degree drops rapidly, reflecting a significant shift in the network's structure. However, at higher values of $b$, the average degree stabilizes, suggesting a saturation point where further increases in $b$ have minimal impact on connectivity. These findings emphasize the crucial role of $b$ in shaping the network's topological characteristics, with different values of $b$ leading to similar trends as the degree values converge at higher $b$ values.

Figure 9 presents the distribution of the shortest path lengths for the constructed networks with values of $b$: $0.01$, $0.1$, $1$, and $100$. For smaller values of $b$, the network exhibits a higher probability of shorter paths, indicating a more interconnected structure conducive to efficient communication. As $b$ increases, particularly for $b = 100$, longer path lengths become more probable, suggesting the network becomes less connected. This trend demonstrates the impact of $b$ on the network's efficiency in terms of node-to-node communication: smaller $b$ values promote efficient traversal, while larger $b$ values lead to longer paths, reducing overall network efficiency.

The relationship between the degree $k$ and the clustering coefficient $C$ across four different values of $b$: $0.01$, $0.1$, $1$, and $100$ is shown in Figure 10. The clustering coefficient decreases as the degree increases, with a more pronounced effect at higher values of $b$. For $b = 0.01$, the clustering coefficient remains high even for larger degrees, indicating a network structure with stronger local connectivity. As $b$ increases to $100$, the clustering coefficient decreases significantly, suggesting the network becomes more random and less cohesive. This finding highlights how $b$ modulates the local interconnectedness of nodes, with lower values fostering tightly connected groups, and higher values promoting a sparser, less cohesive structure.

Figure 11 illustrates the relative size of the giant connected component as a function of the proportion of nodes removed during random attacks. As the proportion of removed nodes increases, the relative size of the giant component decreases for all values of $b$. Networks with lower values of $b$ (e.g., $b = 0.01$) retain a larger giant component even after a substantial portion of nodes is removed, indicating higher resilience to random attacks. Conversely, networks with higher values of $b$ (e.g., $b = 100$) show a more rapid decline in the size of the giant component, suggesting increased vulnerability to random disruptions. These findings underscore the importance of $b$ in determining the network's robustness, with lower values of $b$ enhancing resilience to random failures.

Finally, Figure 12 shows the relative size of the largest connected component as a function of the proportion of nodes deliberately removed during targeted attacks. The data reveal that as the proportion of removed nodes increases, the relative size of the largest connected component decreases for all values of $b$. Networks with lower values of $b$ (e.g., $b = 0.01$) retain a larger proportion of the largest component even after a significant number of nodes are removed, demonstrating higher resilience to targeted attacks. In contrast, networks with higher values of $b$ (e.g., $b = 100$) experience a more rapid decline in the size of the largest component, highlighting increased vulnerability to deliberate disruptions. These observations highlight that networks with lower values of $b$ are more robust against targeted attacks, preserving functionality even when significant portions of the network are compromised.
.
 

\section{CONCLUSIONS}

In summary, we introduce a superstatistical approach to construct $q$-exponential and $q$-Maxwell-Boltzmann complex networks, thereby extending the concept of scale-free networks. By adjusting the crossover parameter $\lambda$, we control the $q$-exponential plateau at low node degrees, allowing a smooth transition to pure power-law degree distributions. Similarly, the parameter $b$ modulates the $q$-Maxwell-Boltzmann curvature, facilitating a shift toward pure power-law networks. This framework provides a novel perspective for constructing and analyzing scale-free networks.Our results demonstrate that these additional degrees of freedom significantly enhance the flexibility of both network types, particularly in terms of their topological and transport properties, such as clustering coefficients\cite{soffer2005network,schank2005approximating}, small-world characteristics\cite{latora2001efficient,klemm2002growing}, and resilience to attacks\cite{klemm2002growing}. Furthermore, this approach offers new insights into the dynamics of network evolution and function.

Future research will focus on exploring the dynamic properties of these networks, including processes such as diffusion, synchronization, and control. By further investigating the impact of the model parameters, we aim to uncover additional network features and contribute to the broader understanding of complex systems. This work opens up promising directions for advancing both theoretical and applied network science.

\section{ACKNOWLEDGEMENT}
I would like to thank my supervisor, weibing deng, for providing the computational resources necessary for this research as well as a place to do the experiment.  I am grateful to tuokui, zhekai chen, and shanshan wang for their assistance and mentorship in this experiment.This work was partially supported by the Fundamental Research Funds for the Central Universities,Natural Science Foundation of China(Grant No.11505071,61702207 and 61873104),and the 111 project 2.0

\bibliographystyle{apsrev}
\bibliography{re}

\end{document}